\documentclass{article}

\usepackage[T2A]{fontenc}
\usepackage[utf8]{inputenc}
\usepackage[english]{babel}
\usepackage{graphicx}

\author{\bfseries Andrey Vasilyev}
\title{\bfseries Distributed Charge in a Coulomb Potential Well}  
\date{Retired from State Optical Institute, Saint Petersburg, Russia
e-mail <andrey@wavemech.org>}

\begin{document}

\maketitle

\begin{abstract}
\label{abstract}

The behavior of  {\itshape\bfseries a distributed} charge in a Coulomb potential well is considered. It is shown that elements of a distributed charge can move along different trajectories, thus forming actually motion of {\itshape\bfseries a charge wave}. Propagation over different trajectories is possible because elements of charge have on these trajectories the same energy. The overall velocity of motion of a charge wave is one half the velocity of individual elements in the wave, and the wave can move only along the $\varphi$ axis of a spherical coordinate system. The following features in the motion of a distributed charge are stressed: each point of the charge is a center from which elements of the charge propagate with the same velocity within the solid angle  $2\pi$. The set of spherical functions needed to describe a distributed charge is specified, and it is shown that a charge cannot have spherical symmetry.
 
\bigskip


\end{abstract}

\section{Motion of a point charge in a Coulomb potential well}

The behavior of a point charge in a Coulomb potential well was studied well
 enough (see, e.g., \cite{1}). By contrast, the behavior of {\itshape\bfseries a distributed} charge in a Coulomb potential well reveals some
 specific features. And it is the specific features arising in the behavior of {\itshape\bfseries a distributed} charge that we are going to address in the present paper. 

We start by assuming a negative distributed charge with a density 
$\rho(\mathbf{r})$ to be placed into the field of a point positive charge $e$ of infinite mass at the origin of a coordinate system. For the sake of conveniency, we shall call this charge a nucleus. Our goal will be to reveal the specific features in the behavior of the distributed charge 
$\rho(\mathbf{r})$ in a Coulomb potential well.
We shall look for the solution of this problem in the framework of the approach that was employed, for instance, in monograph \cite{1}, i.e., in the nonrelativistic approximation. This condition is reliably substantiated.
For instance, in an atom any element of charge located at a distance of the Bohr radius from the nucleus has a velocity on the order of $\alpha c$, where $\alpha$ is the fine structure constant, and $c$ is the velocity of light. Because in the nonrelativistic approximation the velocities of motion of individual charge elements are small, we will disregard in what follows the generation of magnetic field induced by charge motion. Said otherwise, we are going to neglect the vector potential $\mathbf{A}(\mathbf{r})$ induced by charge motion and assume the elements of a distributed charge to feel only the field of the scalar potential $\Phi(\mathbf{r})$.

We shall address in this paper the case where the density of the distributed charge is small. This licenses us to neglect the potential of the distributed charge $\rho(\mathbf{r})$ itself compared with that of the nucleus, and assume each element of the distributed charge to move only in the field of a point charge $e$ with the spherical potential $\Phi(r)=e/r$.

Next we choose in the distributed charge a constant element of charge $dq$ with a mass $dm$ and, assuming this element to be a point charge, recall some features in the behavior of a point charge $dq$ in a Coulomb potential well, which we are going to use later on. (We shall denote this element sometimes by $dq$, and sometimes by $dm$, bearing in mind that in all cases we will deal with an element of charge $dq$ and mass $dm$). A comment is here, however, in order.

An element of charge $dq$ moving with acceleration must radiate energy. Our analysis of the motion of this element in the field of the nucleus will, however, be conducted under the assumption that the element does not radiate. Indeed, the element under consideration, rather than being a single isolated element of charge, has been selected by us out of the total distributed charge. As this will be demonstrated later on, there exist such states of a distributed charge which do not radiate. It is only such states that will be studied in this paper. Therefore in consideration of the motion of a charge element isolated from the total charge we also shall assume that this element does not radiate. Note that each given separate element of charge must radiate.

\medskip

The most convenient approach to identifying the significant details in the behavior of a point charge $dq$ in a Coulomb potential well is to resort to monograph \cite{1}, \$~15. It describes the behavior of a point particle in a field inversely proportional to $r^2$. Among such fields are Newton’s gravitational and Coulomb’s electrostatic fields.
The gravitational forces being weak, we are going to neglect them. The only force we will take account of is the Coulomb interaction of a negative element of charge $dq$ with the positively charged nucleus. Because the element $dq$ separated out from the total charge does not radiate (a point to be substantiated later on), one can invoke here all the conclusions drawn in \cite{1}, \$~15.

Consider now the specific features in the behavior of an element of charge we shall treat in the discussion to follow.

In a Coulomb potential well, a constant element of charge 
$dq$ with a mass $dm$ can move in a circle or an ellipse, 
and the ellipse can degenerate into a straight line. The 
energy of an element of charge depends on the semimajor axis 
of the ellipse (or on the radius of the circle). The actual 
shape of the ellipse (i.e., its semiminor axis) depends on 
the angular momentum of the particle. Thus, all ellipses with 
the same semimajor axis but different semiminor axes have the 
same energies but different angular momenta, down to the zero 
momentum (in which case the ellipse degenerates into a straight 
line). Said otherwise,  elements of the same energy 
can move in a Coulomb potential well along trajectories which 
differ in the value of the angular momentum.

Let us analyze the various trajectories along which an 
element of charge $dq$ with a mass $dm$ can move in the 
case where the total energy of the element in each trajectory 
is the same. 

\begin{figure}
\label{F1}
\begin{center}
\includegraphics[scale=1]{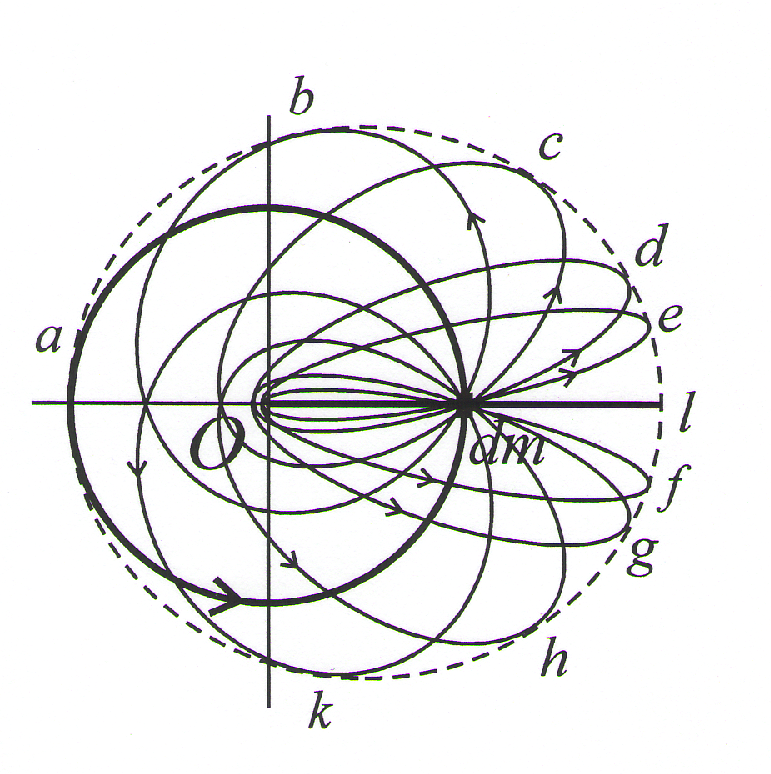}
\caption{}
\end{center}
\end{figure}

Figure~1 for an element $dm$ located at some point
$\mathbf r$ illustrates several such orbits of all possible 
ones: a circular orbit $a$,  eight elliptical orbits 
$(b-k)$ with different eccentricities (and, hence, different 
angular momenta), and a linear orbit $l$ into which the 
ellipse degenerates at an eccentricity of unity. This orbit 
passes through the nucleus. All the orbits are 
characterized by identical semimajor axes (if the energies 
of the elements are equal, the semimajor axes of the ellipses 
should likewise be equal). All orbits lie in the same plane. 
The elements of charge in all orbits rotate in the same 
sense.

All orbits focus at the same point. In this focus (in our 
figure, this is the center $O$ of the circle) the nucleus 
is located. Using the focal properties of 
ellipses, one can readily show that each elliptic trajectory 
intersects a circular orbit at the point where this ellipse 
intersects its semiminor axis. The dashed lines confine the 
region of allowed trajectories along which an element $dm$
can move.

Significantly, the extended trajectories with an eccentricity close to unity pass not far from the nucleus. Therefore, within a small region in the vicinity of the nucleus the condition of the smallness of the element motion velocities does not hold. This region, however, is not large.

Because the energy of an element in each trajectory is the same, element $dm$ can move over {\itshape\bfseries any} of the trajectories specified. As follows from Fig.~1 and an analysis of the trajectories along which element  $dm$ can propagate, all possible directions of motion of the element at a given point are strictly confined to the angle $\pi$ (a few words on a semicircle located above orbit $l$---if the elements rotated in their orbits in the opposite direction, this semicircle would lie below orbit $l$).

An element residing in a Coulomb potential well moves in one plane. Note, however, that the number of planes which can be passed through the line connecting the element under consideration with the nucleus is infinite. Any of these planes can confine the trajectory of the given element, because the Coulomb field possesses spherical symmetry. In this case, all possible directions of the velocities of element motion at a given point lie inside the solid angle $2\pi$ (a hemisphere).
In Fig.~1, this hemisphere extends above the plane passed through orbit $l$ perpendicular to the drawing. If elements in their orbits rotated in the opposite sense, the hemisphere would be located below this plane. Additional conditions related with the position of this plane will be specified below (in Sections 4 and 5).

Consider now the magnitude of the velocities with which 
elements of charge having equal energies move in different 
directions. The Reader can be conveniently referred to monograph 
\cite{1}, \$~14. The energy 
$d\mathcal E$  of an element of charge $dq$ with a mass $dm$ 
moving in a central field is preserved. Recalling the standard 
expression for energy

\begin{equation}
\label{1}
d\mathcal E=d\mathcal E_{Kin}+d\mathcal E_{Pot}=
\frac{dm\upsilon^2}{2}+d\mathcal E_{Pot},
\end{equation}
where $d\mathcal E_{Kin}$ is the kinetic, and $d\mathcal E_{Pot}$, 
the potential energy of the element, and $\upsilon$ is its 
velocity, we obtain

\begin{equation}
\label{2}
\upsilon=\sqrt{\frac{2}{dm}(d\mathcal E-d\mathcal E_{Pot})}.
\end{equation}

This means that the magnitude of the velocity of an element 
with a given energy $d\mathcal E$ depends only on the potential 
energy of this element $d\mathcal E_{Pot}$. Equation (\ref{2}) 
does not contain any other parameters, in particular, of 
parameters which would suggest the dependence of the magnitude 
of the velocity on its direction. As for the potential energy 
of this element, it depends only on the position of this element, 
$d\mathcal E_{Pot}=edq/r$. 
One comes inevitably to the conclusion that the {\itshape\bfseries
{magnitude of the velocity}} of an element at a given point 
{\itshape\bfseries{does not depend on 
the direction}} of motion of this element.

It is a very significant statement, and, hence, it has to be 
corroborated in more detail and in a more revealing way. 
Rewrite Eq. (\ref{1}) in the form

\begin{equation}
\label{3}
d\mathcal E=\frac{dm}{2}(\dot r^2+r^2\dot\zeta^2)+d\mathcal E_{Pot}
=\frac{dm\dot r^2}{2}+\frac{dM^2}{2dmr^2}+d\mathcal E_{Pot}.
\end{equation}
Here $\zeta$ is the angular coordinate in the plane in which 
the element $dm$ rotates, and $dM$ is the angular momentum of 
this element. Whence we come to 

\begin{equation}
\label{4}
\dot r^2=\frac{2}{dm}(d\mathcal E-d\mathcal E_{Pot})-
\frac{dM^2}{dm^2r^2}
\end{equation}
The term $r^2\dot\zeta^2$ can be derived from the expression 
for the angular momentum $dM=dmr^2\dot\zeta$. Substituting it 
into the expression for the velocity, we come to

\begin{equation}
\label{5}
\upsilon=\sqrt{\dot r^2+r^2\dot\zeta^2}=\sqrt{\frac{2}{dm}
(d\mathcal E-d\mathcal E_{Pot})-\frac{dM^2}{dm^2r^2}+
\frac{dM^2}{dm^2r^2}}.
\end{equation}

As seen from Eq.~(\ref{5}), terms containing the angular 
momentum cancel. Said otherwise, the magnitude of the 
velocity does not depend on the angular momentum. Figure~1 
shows, however, that it is different angular momenta of an 
element of a given energy residing at a point that account 
for the different directions of the elements velocity. 

This brings us to the conclusion  that the magnitude of the velocity 
{\itshape\bfseries at a given point} (i.e., at point $\mathbf r$ where element $dm$ resides) does not depend on the trajectory on which element $dm$ moves. At a given point, the element $dm$ has the same velocity, no matter what trajectory is involved.

One more comment is here in order. A separate element of charge $dq$ rotating in a {\itshape\bfseries circular} trajectory radiates an ac field with the frequency of rotation. If, however, elements of the charge fill completely the circular trajectory, this charge distribution will not radiate ac fields, because this state is the steady-state which does not depend on time. The trajectory of such a state is closed and is actually an analog of a circular current. And circular current, as is well known, does not radiate ac fields while generating constant electric and magnetic fields.

\section{Motion of a distributed charge in a Coulomb potential well}

Now recall that we are considering motion not of a point but ruther  
of a distributed charge.

Because in all trajectories which are shown in Fig.~1
the element of charge $dq$ has 
the same energy, it can move 
{\itshape\bfseries along any} trajectory. Moreover, 
this element of charge can move {\itshape\bfseries in all 
trajectories at the same time}. This 
can be visualized in the following way. Divide element   
of charge $dq$ in $k$ parts. Then one element of charge 
$dq'=dq/k$ with mass $dm'=dm/k$ can move along one elliptical 
trajectory, another 
charge $dq'$, along another trajectory, and so on. As $k$ 
tends to infinity, all the trajectories will criss-cross 
all of the allowed region containing trajectories of the 
elements of charge $dq'$ of the same energy but with 
different angular momenta. Generally speaking, this 
process may be considered not as motion of elements of 
charge along trajectories but rather as motion of a 
continuous medium, of a {\itshape\bfseries charge wave}. 

Consider now the velocity with which elements of a distributed charge move. As already shown, elements of a distributed charge can propagate from a given point along all trajectories simultaneously. In Section~1 it was demonstrated that elements of charge have at a specific point the same velocity in any trajectory. At this specific point, these trajectories are characterized by different directions (within a solid angle $2\pi$, see Fig.~1). Hence, when a charge propagates over all trajectories simultaneously, the directions of velocities of different elements at a given point are confined within the solid angle $2\pi$, while their magnitude is the same.

This motion {\itshape\bfseries of a charge wave} may be treated from two angles, to wit, either as motion of different elements of charge on all trajectories at the same time, or as motion of a wave. In the first case, the behavior of each element is described by equations of mechanics (allowing for the potentials in which these elements move; but for the description of the overall picture to be complete, one will have to take into account that the number of these elements is infinite). In the second case, one will have to resort to equations in partial derivatives. For description of the behavior of any continuous medium, and, in particular, of a distributed charge the second model appears to be simpler and, thus, preferable. For the present, however, we are going to adhere to the first approach -- it appears more graphic (while certainly more cumbersome). 

Turning now to Fig.~1, we see immediately that different trajectories (i.e., trajectories characterized by the same energy but different momenta) cross one another. Said otherwise, it turns out that the charges moving along all trajectories simultaneously {\itshape\bfseries can “interpenetrate”} one another without  changing their trajectories.

This would seem at first glance to be in contradiction with the well known statement that like charges repel. Therefore the assumption that likely charged elements can penetrate into one another may sound surprising, to say the least. But point charges considered usually in science have an infinite density at the point of charge. Therefore like point charges cannot penetrate into one another  and, moreover, cannot even approach one another to a short enough distance. The distributed charges considered by us have a specific finite charge density. We are going to show now that charges can penetrate into one another, depending on what external forces act on these charges and what relevant forces are generated by the charges themselves.

The force $d\mathbf F=\mathbf Edq$ acting on any element of charge 
$dq=\rho dV$ is defined by the magnitude of the electric field 
$\mathbf E$ at this point. The field $\mathbf E(\mathbf r)$ at a given point  $\mathbf r$ is a sum of the field $\mathbf E_N(\mathbf r)$ generated by the nucleus and the field $\mathbf E_{\rho}(\mathbf r)$ created by all the charges surrounding the given charge:

\begin{equation}
\label{6}
\mathbf E(\mathbf{r})=\mathbf E_N(\mathbf{r})+\mathbf E_{\rho}(\mathbf{r})=
\frac{e\mathbf r}{r^3}+\int{\frac{\rho(\mathbf r')(\mathbf r-\mathbf r')dV'}
{|\mathbf r-\mathbf r'|^3}}.
\end{equation}

The presence of expression $|\mathbf r-\mathbf r'|^3$ in the denominator of the second term in equality (\ref{6}) might cause an erroneous idea that the field $\mathbf E_{\rho}(\mathbf{r})$ at point $\mathbf r$ is large. In this case the element of charge located at point $\mathbf r''$, near point
 $\mathbf r$, will not be able to approach in its motion point 
$\mathbf r$, because both these elements of charge have like signs.
 One can readily show that this is not so: indeed, the field 
$\mathbf E_{\rho}(\mathbf{r})$ is finite at point $\mathbf r$, and the difference between the fields at the two neighboring points $\mathbf r$ and   $\mathbf r''$ tends to zero as the magnitude of 
$|\mathbf r-\mathbf r''|$ approaches zero. 

To prove this, we use the approach employed in monograph \cite{2}, $\$44$. Circumscribe a sphere of radius $R_0$ around point $\mathbf r$. The field generated by the charges outside the sphere $R_0$ is finite, because these charges are at a finite distance larger than $R_0$ from point $\mathbf r$. We have now to verify that the field $\hat\mathbf E(\mathbf r)$ generated by charges confined inside the sphere $R_0$, i.e., in the immediate vicinity of point $\mathbf r$, is also finite. Denoting  $|\mathbf r-\mathbf r'|=R$, and 
$(\mathbf r-\mathbf r')=\mathbf R$, we can write for this part of the field:

\begin{equation}
\label{7}
| \hat \mathbf E(\mathbf r)|\le\int{\frac{|\rho(\mathbf r')\mathbf R|d\hat V}
{R^3}},
\end{equation}
where integration of $d\hat V$ is performed over the volume of the sphere 
$R_0$. But
$$
|\rho(\mathbf r')\mathbf R|\le |\rho_{max}|R,
$$          
where $|\rho_{max}|$ is the absolute value of the maximum density of charge inside sphere $R_0$.

We finally come to 
$$
| \hat \mathbf E(\mathbf r)|\le|\rho_{max}|\int{\frac{d\hat V}
{R^2}}.
$$    

Introducing spherical coordinates  $R, \vartheta, \varphi$ with the center at point $\mathbf r$, with $d\hat V=R^2sin\vartheta d\vartheta d\varphi dR$,
and integrating with respect to $R$ from $0$ to $R_0$, we obtain

\begin{equation}
\label{8}
|\hat \mathbf E(\mathbf r)|\le 4\pi|\rho_{max}|R_0.
\end{equation}

Thus $\hat \mathbf E(\mathbf r)$ is a finite quantity tending to zero with decreasing radius of the sphere $R_0$. Moreover, this immediately suggests a conclusion that the difference between the values of vector $\mathbf E$ 
at two adjacent points, for instance, $\mathbf r$ and $\mathbf r''$, tends to zero with the distance between these points approaches zero too. Suppose that these two points are located inside the sphere $R_0$. The field created by charges outside the sphere $R_0$ is continuous, because these charges are at finite distances from points $\mathbf r$ and $\mathbf r''$.
 As for the field $\hat \mathbf E$ generated by charges confined inside the sphere $R_0$, the strength of this field in absolute magnitude, as proved above, cannot be larger than the value of $4\pi|\rho_{max}|R_0$. As the magnitude of $R_0$ is going to zero, we see that this part of the field also changes continuously and tends to zero as $R_0$ approaches zero.

Thus the electric field $\mathbf E$ surrounding any element of the distributed charge $dq=\rho dV$ is finite and varies continuously. And this is why the force $d\mathbf F=\mathbf Edq$ acting on this element of charge is finite and varies continuously. The element of distributed charge will be driven by this force to move in the direction of the total force acting at this point.  

In actual fact, the statement that charges can pass through 
one another does not carry anything supernatural in it. For 
instance, electromagnetic fields can penetrate one into or 
through the other without at the same time affecting one 
another---this is nothing but the standard principle of 
superposition. Two radar beams can cross without interaction; 
this is just penetration of ac fields through one another. 
Superposition of one dc field on another (the principle of  
superposition) may be regarded as penetration of one field 
into another. Significantly, in this process the fields do 
not act in any way on one another.

As for the charges, no statements concerning passage of one 
charge through another without direct action on one another 
(interaction of charges is taken into account through the 
fields created by these charges) have thus far been made, 
although the principle of superposition is valid for charges 
as well. This statement should, however, be made. 

This paper was intended to address the case of small density of the distributed charge. This means that we shall neglect the second term in expression (\ref{6}) compared to the first one. In this case, any element of charge $dq$ with mass $dm$ moves only in the field of charge of the nucleus 
$e$, and for this element all the conditions specified in Section~1 are fully met, and the motion of this element will be subject to the laws described in the above monograph \cite{1}, \$~15.

We have considered earlier a set of trajectories, both elliptical and circular, which are characterized by the same energy. In a Coulomb potential well, however, elements moving along different, including circular, trajectories may have different energies. The potential energy of an element in a circular trajectory is constant and depends only on the distance of this element from the nucleus, $d\mathcal E_{Pot}=edq/r$. Each element in any circular trajectory can be identified by a set of elliptical trajectories with the same energy (see Fig.~1). This makes the total set of all trajectories extremely complex.

\section{Distributed charge has no spherical symmetry}

Consider the shape which can have a charge in a Coulomb potential well.

In a spherical coordinate system $r, \vartheta, \varphi$ the angular part of any distribution of charges can be presented in the form of an expansion in spherical functions $Y_{lm}(\vartheta, \varphi)$. Let us see what spherical functions can be employed in description of a distributed charge in a Coulomb potential well.

The simplest spherical function is the spherically symmetric function 
$Y_{00}(\vartheta, \varphi)$. This function should be present in an expansion always (except for the cases where the total charge of the distribution is zero). This has the following natural explanation. The integral over all space of a charge expanded in spherical functions $Y_{lm}(\vartheta, \varphi)$ yields the total charge $Q$. But the only angular function whose integration yields a nonzero result is $Y_{00}(\vartheta, \varphi)$. Integration of all other functions will yield zero. 
Therefore only the function $Y_{00}(\vartheta, \varphi)$ can describe the presence itself of a charge in a volume. All the other angular functions participating in the expansion can only change the shape of the charge distribution, while not being capable of removing or adding a charge.

On the other hand, a distributed charge cannot be characterized with the use of one angular function $Y_{00}(\vartheta, \varphi)$ only. This becomes evident from the following consideration. A Coulomb potential well is spherically symmetric. Therefore, an elliptical trajectory of propagation of an element of charge may lie in any plane passing through the nucleus (i.e., through the origin of coordinates).

An analysis of all elements of a distributed charge reveals that their trajectories lie {\itshape\bfseries in all planes simultaneously}. Consider now different planes in which the elliptical trajectories of elements may lie. We choose for this purpose one of such planes, rotate it successively and follow several trajectories of motion of an element. The continuity of the distributed charge allows this operation.

\begin{figure}
\label{F2}
\begin{center}
\includegraphics[scale=1]{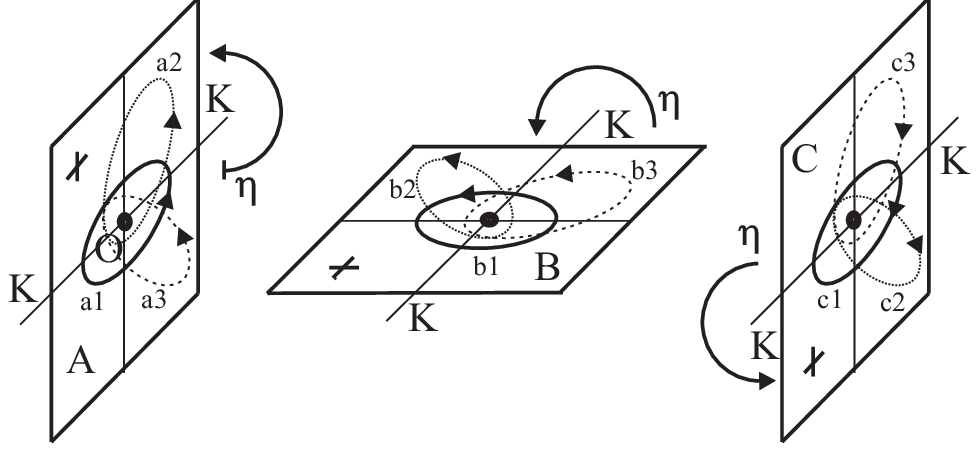}
\caption{}
\end{center}
\end{figure}

This situation is visualized in Fig.~2. It shows three planes which may confine the trajectories of motion $A$, $B$ and $C$. The nucleus is located at point $O$. A $K-K$ line passing through the nucleus is drawn in plane $A$. One can see three trajectories in plane $A$: a circular trajectory $a1$, an elliptical trajectory $a2$, and another elliptical trajectory $a3$, which is symmetric to $a2$ about the $K-K$ line.
It is shown that all elements lying in plane $A$ rotate in the same sense. We shall turn plane $A$ about the $K-K$ line in angle $\eta$ starting from an initial direction. Because the Coulomb potential is spherically symmetric, and the charge continuous, it can be expected that on turning plane $A$ about the $K-K$ line through angle $\pi$, we shall come to the state in which all trajectories will coincide with the ones that had been there before the turn of the plane.

Let us perform this operation of the turn. Figure~2 shows two positions of the plane after a turn by $\pi/2$ and by an angle $\pi$. By turning plane $A$ about the $K-K$ axis in angle $\eta$ by an angle $\pi/2$, we come to plane 
$B$ with trajectories $b1$, $b2$, $b3$, and by angle $\pi$, to plane $C$ with trajectories  $c1$, $c2$, $c3$. 
Significantly, trajectory $c1$ in plane  $C$ will exactly coincide with trajectory $a1$ in plane $A$. Similarly, trajectory  $a2$ will merge with trajectory $c3$, and trajectory  $a3$, with $c2$. Thus for each elliptical trajectory in plane $A$ there will be always the corresponding trajectory in plane $C$. After the turn of the plane, these trajectories will coincide.

It will turn out, however, that motion along all these trajectories after completion of the turn of the plane (i.e., in position $C$) will occur in the sense opposite to that in which the elements moved in the plane before its turn (that is, in position $A$). The + sign on planes $A$, $B$, $C$ is put for the sake of convenience in following the effect of plane rotation.

Since we are considering a {\itshape\bfseries distributed charge}, its elements should move simultaneously in all planes, including planes $A$ and $C$. As showed, however, our analysis, motion over the latter should occur in opposite directions. All elements just cannot rotate simultaneously in opposite directions (this would involve loss of energy).

It thus turns out that motion over these trajectories is impossible altogether. Hence, there should exist a direction in which trajectories for motion of charges do not exist at all. In the spherically symmetric potential well of the nucleus one cannot, however, isolate a specified direction (for instance, the direction from which we reckoned the angle  $\eta$ in Fig.~2). Hence, a distributed charge must have a specified direction, a direction in which the distributed charge does not exist. In this case, with no charge in this direction, one cannot expect existence of trajectories along which propagation could occur in opposite directions.

In other words, although the Coulomb potential is spherically symmetric, distributed charge loses spherical symmetry.

This appears only natural, because the angular momentums  in plane $A$ (i.e., before the turn of the plane) do not coincide with those in plane $C$ (i.e., after the turn of the plane) and, moreover, have opposite orientation.

To sum up, in a spherically symmetric Coulomb potential well a spherically symmetric charge distribution just cannot exist; there must therefore exist a specified direction. An additional comment concerning a specified direction in which there is no distributed charge will be proposed in Section 5.

\section{Description of a charge with spherical functions}

As follows from previous considerations, description of a distributed charge has to be made with a spherically symmetric angular function 
$Y_{00}(\vartheta, \varphi)$ (this function specifies the presence itself  of a charge in the given potential well). This function alone is not sufficient, however, because the $Y_{00}(\vartheta, \varphi)$ function does not have any specified direction (a specified direction appears as a result of the circular motion of the elements of charge around the nucleus). Therefore, description of a charge requires, even in the simplest case, invoking some other $Y_{lm}(\vartheta, \varphi)$ functions as well. Let us see what functions could be employed in description of a distributed charge.

We start by assuming that the specified direction discussed above coincides with the $\vartheta=0$ direction of the spherical coordinate system. We shall call this direction the $Z$ axis. Then total rotation of the elements of charge will occur about the $Z$ axis, i.e., along the $\varphi$ coordinate. We understand under total rotation here not the rotation of any one element but rather that of the totality of the elements, of the distributed charge as a whole. We discussed in Section 1 a plane with respect to which an element   $dq$ (or $dm$) propagates as a wave in a solid angle $2\pi$. In the present case this plane passes through the axis $Z$ and the position of the element at the given moment, i.e., perpendicular to the $\varphi$ coordinate.

Introduce an additional condition; to wit, we are going to consider only stationary, i.e., time-independent, states of the charge. States of charge which do not depend on time, do not produce radiation of variable fields, while generating constant electric and magnetic fields. Because stationary states of a distributed charge do not vary with time, no need appears in proving that such states do not generate variable fields, i.e., fields depending on time. If total rotation of the charge occurs about the $Z$ axis, i.e., along the $\varphi$ coordinate, absence  of radiation can be described by using for description  only functions which are axially symmetric about the $Z$ axis. This restricts the allowable set of functions to those of the $Y_{l0}(\vartheta, \varphi)$ group. Indeed, only 
$Y_{l0}(\vartheta, \varphi)$ functions do not depend on the coordinate 
$\varphi$, i.e., have axial symmetry with respect to the $Z$ axis.

If we impose one more constraint, namely, that the distributed charge is symmetric with respect to the coordinate $\vartheta=\pi/2$, i.e., about the equator (which is a more frequent situation), index $l$ of the spherical function $Y_{l0}$ can be only even.

There are no other proper functions for description of a charge which does not generate radiation. Indeed, $Y_{lm}(\vartheta, \varphi)$ functions contain a factor $exp(\pm im\varphi)$. Motion of such a charge along the coordinate $\varphi$ will initiate dependence on time (a factor of the kind of $exp(\pm im\varphi-i\omega t)$ will appear, where $\omega$ is the frequency of the moving wave). The appearance of the dependence on time will inevitably give rise to radiation of variable fields. We disregard here such states involving radiation and focus our interest on stationary states only, which do not generate radiation.

The simplest function satisfying the above requirements is $Y_{20}$. We write therefore the angular part of the relation for a distributed charge in the form

\begin{equation}
\label{9}
L=D(Y_{00}+D_{20}Y_{20}),
\end{equation}
$$
\mbox{where  }Y_{00}=\frac{1}{\sqrt{4\pi}},\quad Y_{20}
=\sqrt{\frac{5}{4\pi}}\left(\frac{3}{2}\cos^2\vartheta-\frac{1}
{2}\right),\quad \mbox{(see, e.g., \cite{3}).}
$$

The coefficient $D_{20}$ can be determined from the condition that at 
$\vartheta=0$ and $\vartheta=\pi$ (i.e., on the $Z$ axis) the function $L$ be zero. Coefficient  $D$ can be found from the condition that at 
$\vartheta=\pi/2$ (i.e., at the equator) function $L$ is unity. These conditions bring us to

$$
D=\frac{2\sqrt{4\pi}}{3},\qquad D_{20}=-\frac{1}{\sqrt{5}}.
$$
Now function L acquires the final form
$$
L=\frac{2\sqrt{4\pi}}{3}\left[\frac{1}{\sqrt{4\pi}}-\frac{1}
{\sqrt5}\cdot\sqrt{\frac{5}{4\pi}}\left(\frac{3}{2}\cos^2
\vartheta-\frac{1}{2}\right)\right].
$$
One can readily verify that this function simply coincides with the function 
$\sin^2\vartheta$. This means that in this case the angular part of the density of distributed charge can be written in one of two ways:

\begin{equation}
\label{10}
L=\sin^2\vartheta, \quad \mbox{or} \quad L=D(Y_{00}+D_{20}Y_{20}),
\end{equation}
and the density of distributed charge corresponding to this angular distribution will read
\begin{equation}
\label{11}
\rho(r,\vartheta)=AR(r)\sin^2\vartheta, \quad \mbox{or} \quad
\rho(r,\vartheta)=AR(r)D(Y_{00}+D_{20}Y_{20}),
\end{equation}
where $R(r)$ is the radial part of the distribution. The coefficient $A$ is derived from normalization of the distributed charge against the total charge $Q$.

One may choose any form that would seem appropriate in a given situation.

Thus, in describing a charge with spherical functions one obtains for the charge distribution in a Coulomb potential well in the simplest case a figure resembling a torus. All elliptical and circular trajectories of an elements of the charge (of any energy) should be confined to this torus. We note that at the $Z$ axis the charge is zero.

A comment will be appropriate here. As shown earlier, at a given point the element $dm$ propagates in all directions (within a solid angle $2\pi$) with the same velocity. It would seem that this is in direct contradiction with the statement that a distributed charge has a specified direction in which there is no charge. In particular, in the above example with a torus-shaped charge, it would seem that the trajectories lying on the equator must differ from those confined to a perpendicular plane, because these trajectories cross the $Z$ axis, where the charge is zero. 

The following point may be in order here. The trajectories of the element $dm$ in Fig.~1 were considered by us under the tacit assumption that this element will continue to move along its original trajectory. (This assumption derives from the concept of the motion of a solid body.) This is, however, not necessarily so. An analysis of Fig.~1 shows that {\itshape\bfseries each point} is a center from which elements of charge propagate in all directions (within a solid angle $2\pi$). Said otherwise, element $dm$ does not move along this trajectory all the time. At each point it breaks up into many elements $dm'$ which continue to move subsequently, but now along other trajectories.
 At the next point the situation repeats, with breakup into many elements, at the next point -- again into countless elements, and so on. Thus, the element $dm$, in starting its motion at a point on a trajectory, should not necessarily terminate it at the same trajectory. Actually, this is motion not of individual elements but rather that of a wave. This is why a mass (and a charge) may have different densities at different points in space.

This is an allowed process, because in each trajectory the corresponding element has the same energy. But {\itshape\bfseries at each given point} the velocity of the elements remains, as before, the same in all directions, irrespective of the density of charge or mass at the given point.

\section{Differences in the velocity of motion between a charge wave and individual elements}

As already pointed out, it appears more appropriate to consider the motion of elements of equal energy along different elliptical trajectories as that of a charge wave. It appears pertinent to compare now different parameters of motion of this wave with those of an individual element.

The total energy of all elements making up a charge wave is equal to the energy of one combined element lying at the point of crossing of all elliptical trajectories and moving along one of the circular trajectories. Considered in the context of equality of energies, the motion of a charge wave may be correlated with that of one point element, with the sum of the energies of all elements in this wave equal to the energy of one combined element moving in a circle. This energy can be readily determined.

An element moving along a circular trajectory retains both its total and the potential and kinetic energies. By the virial theorem (see, e.g. \cite{1}, 
\$~10), in this case the potential energy of an element is twice its total energy, and its total energy is equal to the kinetic energy taken with the opposite sign, 
$2d\mathcal E=d\mathcal E_{Pot}$, $d\mathcal E=-d\mathcal E_{Kin}$. No averaging is needed here, because the energies are constant. As for the potential energy of an element in a circular orbit, it can be derived simply from the radius $R$ of the circular orbit: $d\mathcal E_{Pot}=edq/R$. 

The situation is different with the velocities of motion of a wave and of individual elements.
 
An analysis of Fig.~1 suggests that {\itshape\bfseries each point} of a charge is a center from which elements of charge propagate in all directions (within a solid angle $2\pi$). One might say that the motion of a distributed charge at a given point represents a kind of “a velocity fan”\ for all directions (within a solid angle $2\pi$); note that, as shown in Section~1, in any direction the velocity has the same magnitude. 

The velocity of an element being a vector, the velocity must 
retain its vector properties even in the case of the element 
propagating (in the form of a wave) within a solid angle of 
$2\pi$. This, however, will be not the velocity of a single 
element but rather that of motion of a wave, of propagation 
of a wave process. It turns out that the velocity of propagation 
of a wave process does not coincide with that of motion of 
elements of mass or charge. The momentum of an element $dm$ 
propagating as a wave likewise does not coincide with that 
of an element moving as a whole in one direction.

Denote the velocity of motion of a charge wave by $\mathbf 
v_w(\mathbf r)$ (the subscript $w$ standing for wave), and 
the magnitude of this velocity, by $\upsilon_w(\mathbf r)$, 
to discriminate this velocity from the velocity 
$\mathbf v(\mathbf r)$ and $\upsilon(\mathbf r)$ of motion 
of an element of charge. The momentum of an element $dm$ 
propagating as a wave will be denoted, accordingly, by 
$\mathbf p_w(\mathbf r)$.

Consider this situation in more detail.

Isolate an element of mass $dm$ at a point in space. If this 
element moves as a whole with a velocity $\mathbf{v}_l$ along 
a trajectory $l$, the momentum $d\mathbf{p}_l$ of this element 
will be
\begin{equation}
\label{12}
d\mathbf{p}_l=dm\mathbf{v}_l,\qquad\mbox{and the magnitude of 
the momentum }\qquad dp_l=dm\upsilon.
\end{equation}

To describe the motion of this element of mass as that of a wave 
in the solid angle of $2\pi$, divide the mass $dm$ into many parts 
$dm'$. Each part $dm'$ will propagate within a solid angle $d\Omega$. 
In this case, we can write $dm'=dm\cdot\frac{d\Omega}{2\pi}$. 
As was shown in Section~1
the velocity of motion of each element $dm'$ is the same, equal 
in magnitude to the velocity $\upsilon$ of motion of the whole 
element $dm$. 
 The direction of this velocity 
$\mathbf{v'}$ is determined by the solid angle $d\Omega$. 
For the momentum of element $dm'$ in this case we can write
\begin{equation}
\label{13}
d\mathbf p'=dm'\mathbf v'=dm\frac{d\Omega}{2\pi}
\mathbf v',\quad\mbox{and for its magnitude}\quad
dp'=dm\frac{d\Omega}{2\pi}\upsilon.
\end{equation}

It might come up as a surprise that sometimes we discuss motion 
along an elliptical trajectory (for instance, trajectory $l$) 
to stress that the element $dm$ moving in this trajectory obeys 
all laws of theoretical mechanics, while in other cases we prefer 
to identify the motion of the same element in a solid angle as 
that of a wave.

In actual fact, we are speaking in these cases about different 
things. When discussing the motion along a trajectory, we follow 
the motion of {\itshape\bfseries{one}} specific element, be it 
element $dm$ or $dm'$. It appears only natural that the motion 
of this element obeys all laws of theoretical mechanics.

When, however, we discuss the motion of an element as a wave in 
a solid angle, we have in mind rather the motion of 
{\itshape\bfseries{many}} elements crossing at a point 
$\mathbf r$. This is shown in a revealing way in Fig.~1. 
This set off elements could be formed of one element $dm$ as well. 
It is for this purpose that we broke up element $dm$ into a set 
of elements $dm'$, which thereafter moved along {\bfseries\itshape
{different}} trajectories.

\medskip

We chose to orient a spherical coordinate system such that the 
direction $\vartheta=0$ coincides with that of the angular momentum 
of the charge, and denoted this axis by $Z$. All elements of a 
distributed charge rotate in the same sense (overall rotation 
is around the $Z$ axis). That all elements rotate in one sense only 
implies that there are no velocity components along the negative 
direction of the $\varphi$ axis. In other words, on passing a 
plane through the $Z$ axis and the position of element $dm$, we end 
up with the following situation: element $dm$ propagates (as a wave) 
into a solid angle of $2\pi$, i.e., into the hemisphere located 
on one side of this plane.

\begin{figure}
\label{F3}
\begin{center}
\includegraphics[scale=1]{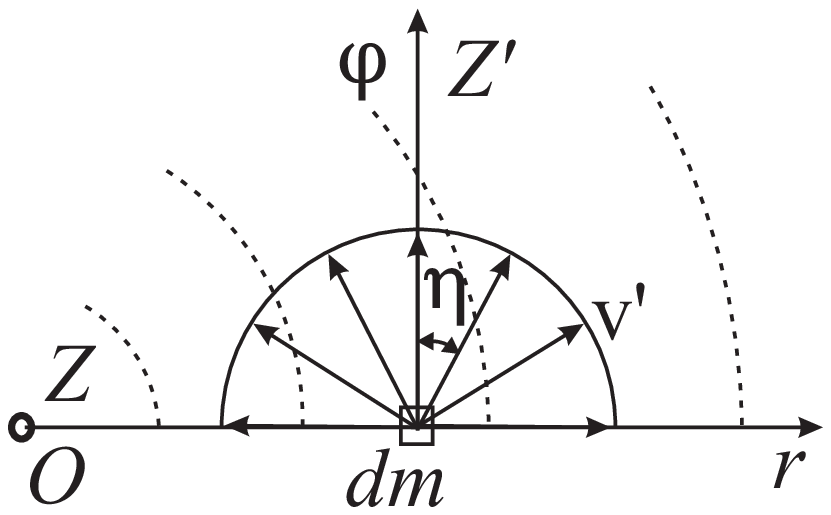}
\caption{}
\end{center}
\end{figure}

This situation is illustrated by Fig.~3. In Fig.~3, the $Z$ axis is 
directed at us (i.e., it is actually a top view). The nucleus  
is at point O. Dashed lines are lines of equal mass density. 
Element $dm$ propagates into a solid angle of $2\pi$, i.e., into the 
hemisphere. We see a fanlike distribution of velocities $\mathbf v'$ 
of the elements of mass $dm'$. The velocity of motion of element 
$dm'$ in any direction is the same. In Fig.~3, this is shown by all 
velocity vectors $\mathbf v'$ being of equal length.

To find the resultant momentum of an element $dm$ in the case where 
it propagates (as a wave) into a hemisphere, we have to sum all the 
momentum vectors $d\mathbf p'=dm'\mathbf v'$. This can be done by 
introducing a local spherical system of coordinates centered on the 
location of element $dm$, with angles $\eta$ and $\xi$. We orient 
the coordinate system such that the $\eta=0$ direction is 
perpendicular to the above-mentioned plane and denote this direction 
by $Z'$. In this particular case, $Z'$ coincides with the direction 
of the $\varphi$ axis of the common coordinate system. Next we 
resolve the momentum into a component along the $Z'$ axis, and another,
perpendicular to it. By virtue of the symmetry relative to 
the $Z'$ axis, the perpendicular component will vanish after the 
summation, leaving only the component along the $Z'$ axis, i.e., 
along the $\varphi$ axis. Thus, the resultant momentum of an element 
moving in all directions simultaneously (into a solid angle of $2\pi$, 
i.e., into the hemisphere) is aligned with the $\varphi$ axis. 
Calculate now $dp_{w,\varphi}$, i.e., the projection of the momentum 
on the $\varphi$ axis. To do this, we sum all $dp'_\varphi$ components. 
Using Eq. (\ref{13}), we come to $dp'_\varphi=dp'\cos\eta=
dm'\upsilon\cos\eta=dm\frac{d\Omega}{2\pi}\upsilon\cos\eta$, whence:
\begin{equation}
\label{14}
dp_{w,\varphi}=\int{dm\frac{d\Omega}{2\pi}\upsilon\cos\eta}=
\frac{dm}{2\pi}\upsilon\int\limits_0^{\pi/2}\cos\eta
\sin\eta d\eta\int\limits_0^{2\pi}d\xi=
\frac{1}{2}dm\upsilon=\frac{1}{2}dp_l,
\end{equation}
or
\begin{equation}
\label{15}
d\mathbf p_w=\frac{1}{2}dm\upsilon \mathbf n_\varphi ,
\end{equation}
where $\mathbf n_\varphi$ is the unit vector along the $\varphi$ axis.

Comparing now Eqs. (\ref{14}) and (\ref{15}) with relations (\ref{12}), 
we see that the momentum of an element propagating into a hemisphere 
is one half that in magnitude of an identical element moving in one 
direction. Also, while the momentum of an element moving as a whole 
in one direction coincides in direction with its velocity, the 
momentum of an element propagating into a hemisphere is directed 
{\itshape\bfseries {only along the}} $\varphi$ {\itshape\bfseries 
{axis}}. {\itshape\bfseries {This momentum has no other components}}.

The above reasoning and the conclusions were conducted for a momentum. 
To obtain a diverging  wave, the element $dm$ was broken up 
into smaller elements $dm'$, with all $dm'$ elements having the same 
velocity $\upsilon$, and each element of mass $dm'$ propagating in 
its solid angle $d\Omega$. This is a rigorous treatment. It can be 
made more convenient, however, by considering not the momentum but 
rather directly the velocity. To do this, we denote conventionally 
 $d\upsilon=\upsilon d\Omega$, understanding by $d\upsilon$  a set 
 of velocities with directions confined to the solid angle $d\Omega$. 
 Then in place of Eq. (\ref{15}) and taking into account (\ref{12})
 we come to
\begin{equation}
\label{16}
\mathbf v_w=\frac{1}{2}\upsilon\mathbf n_\varphi.
\end{equation}

Here $\mathbf v_w$ is no longer the velocity of a single element 
$dm$ moving in its trajectory; it is now the remaining vector part 
of the velocity of the element $dm$ propagating into the hemisphere, 
and $\upsilon$ is, as before, the magnitude of the velocity of the 
element moving in its trajectory.  
(Equation (\ref{16}) can be also derived 
directly from Eq. (\ref{15}) by canceling $dm$).

Thus, the velocity $\mathbf v_w(\mathbf r)$ is the velocity of 
propagation of a wave process, and the modulus of 
$\mathbf v_w(\mathbf r)$ is the magnitude of the velocity of the 
wave process $\upsilon_w(\mathbf r)$. The magnitude of the velocity 
of a wave process does not coincide with that of propagation of 
elements of charge. Indeed, taking the modulus of $\mathbf v_w$ 
(using Eq. (\ref{16}) and recalling that vector $\mathbf v_w$ has 
only one component, and it is directed along the $\varphi$ axis), 
we obtain only one half of the real velocity of an element $\upsilon$. 
Indeed, some of the vector components vanish in propagation of 
the element as a wave into the hemisphere. 
Actually, these components do not disappear without trace, so that, 
for instance, they have to be taken into account in calculation of 
the energy, because in actual fact elements move along elliptical 
trajectories with a velocity $\upsilon$.
Specifically, for calculation of the kinetic energy one must use
the $\upsilon$ quantity that is the real velocity of an element.
For calculation of the angular momentum (vector quantity) 
of the distributed charge one must use $\mathbf v_w$ quantity.

\end{document}